\documentclass[doublecol]{epl2} 

\begin{document}

\title{Rapidly-oscillating scatteringless non-Hermitian potentials and the absence of Kapitza stabilization}
\shorttitle{Rapidly-oscillating scatteringless  ... } 

\author{S. Longhi \inst{1,2}}
\shortauthor{S. Longhi}

\institute{                    
  \inst{1}  Dipartimento di Fisica, Politecnico di Milano, Piazza L. da Vinci 32, I-20133 Milano, Italy\\
  \inst{2}  Istituto di Fotonica e Nanotecnlogie del Consiglio Nazionale delle Ricerche, sezione di Milano, Piazza L. da Vinci 32, I-20133 Milano, Italy}
\pacs{03.65.Nk}{Scattering theory}
\pacs{03.65.-w}{Quantum mechanics}

\abstract{In the framework of the ordinary non-relativistic quantum mechanics,  it is known that a quantum particle in a rapidly-oscillating bound potential with vanishing time average can be scattered off or even trapped owing to the phenomenon of dynamical (Kapitza) stabilization. \textcolor{red}{A similar phenomenon occurs for scattering and trapping of optical waves}. Such a remarkable result stems from the fact that, even though the particle is not able to follow the rapid external oscillations of the potential, these  are still able to affect the average dynamics by means of an effective -albeit small- nonvanishing potential contribution. Here we consider the scattering and dynamical stabilization problem for matter or classical waves by a bound potential with oscillating ac amplitude $f(t)$ in the framework of a non-Hermitian extension of the Schr\"odinger equation, and predict that for a wide class of {\it imaginary} amplitude modulations $f(t)$ possessing a one-sided Fourier spectrum the oscillating potential is effectively canceled, i.e. it does not have any effect to the particle dynamics, contrary to what happens in the Hermitian case.}
\maketitle

\section{Introduction}

Periodically-driven classical and quantum systems are known to show rich dynamical behaviors of major relevance in different areas of physics   \cite{r1,r2}. 
Examples include dynamical localization in the periodically-kicked quantum rotor model \cite{r3,r4}, adiabatic stabilization of atoms irradiated by ultra strong electromagnetic fields in the high-frequency regime \cite{r5,r6,r7}, coherent destruction of tunneling and dynamic localization \cite{r1,r8,r9,r10,r11,r11bis}, Kapitza stabilization in rapidly-oscillating potentials \cite{r12,r13}, field-induced barrier transparency \cite{r14}, driven quantum motors \cite{r15}, Floquet engineering and synthetic phases of matters \cite{r16,r17,r18,r19,r20,r21}, etc.  A common feature underlying the rich dynamical features found in driven systems  is the fact that the high-frequency limit of the Floquet Hamiltonian of the periodically-driven system is not
trivially equal to the time-averaged Hamiltonian, so that a rapid modulation can be exploited to tailor the system dynamics. A paradigmatic example is provided by the dynamical (Kapitza) stabilization effect, i.e. the possibility for a classical or quantum particle to be trapped by a rapidly oscillating potential in cases where the static potential cannot trap them \cite{r2,r12,r13,r22,r23,r24,r25,r26,r27}. The prototypical
example of dynamical stabilization is the Kapitza pendulum, i.e. stabilization of an inverted pendulum when the point of suspension is being displaced periodically along the vertical direction \cite{r12,r13}. The original idea of Kapitza pendulum has been successfully extended into quantum \cite{r22,r23,r24,r25,r26,r27,r27bis,r27tris} and nonlinear  \cite{r28,r29,r30} physics, with important applications in Paul traps for charged particles \cite{r31}, in driven
bosonic Josephson junctions \cite{r32},  in nonlinear optical dispersion
management \cite{r28,r30}, and in light guiding and diffraction control in optics \cite{r33,r34,r35,r36,r37}.  \textcolor{red}{We note that Kapitza stabilization can be found for ac oscillations that are periodic, quasi periodic or even stochastic}. Several other facets of dynamical stabilization have been investigated, such as the role of the potential phase \cite{r38,r39,r40,r41} and the extension of the Kapitza stabilization to an imaginary potential \cite{r42}. 
 In the framework of the quantum dynamics of a particle in a rapidly oscillating bound potential \cite{r22,r25,r26,r27,r27bis}, Kapitza stabilization arises from the fact that, even though the particle is not able to follow the rapid external oscillations and the time-averaged potential vanishes, the potential oscillations are still able to affect the average dynamics by means of an additional -albeit small- potential contribution, which can sustain quasi-bound (resonance) modes \cite{r27}. An intriguing question is whether one can find cases where the rapidly oscillating potential does not provide any effect on the particle dynamics, corresponding to the absence of Kapitza stabilization and particle scattering. Such a kind of potentials are prevented in ordinary \textcolor{red}{Hermitian systems}, since the residual effective potential does not vanish \cite{r27}. Several recent works have suggested that scattering and localization properties of classical or matter waves are deeply modified \textcolor{red}{in systems described by an effective non-Hermitian models with complex potentials \cite{r43,r44,r45,r46,r47,r48,r49,r50,r50bis,r50tris,r51uff,r51,r52,r53,r54}. Non-Hermitian Hamiltonians, especially those  possessing parity-time symmetry \cite{r55}, have received a huge and increasing attention in recent years, revealing a wealth of interesting effects with rapidly emerging applications, especially in photonics. For example, non-Hermitian Hamiltonians supporting exceptional points show a rich variety of intriguing phenomena such as loss-induced transparency, band merging, unidirectional invisibility, asymmetric encircling dynamics, topological energy transfer, etc. Impressive implementations of engineered non-Hermitian Hamiltonians have been reported in several recent experiments, including optical \cite{  r47,r48,r52,r53}, optomechanic \cite{r54}, and matter wave \cite{r50bis,r50tris} systems}. However, the dynamical aspects of non-Hermitian rapidly oscillating potentials have received few attention so far from the point of view of localization and scattering features \cite{r42,r51}.\\
  In this Letter we consider the scattering and dynamical stabilization properties of \textcolor{red}{classical or matter waves} in a bound potential $V(x)$ with an ac oscillating complex amplitude $f(t)$, and predict that for a wide class of {\it imaginary} amplitude modulations $f(t)$ possessing a one-sided Fourier spectrum the oscillating potential is effectively washed out, i.e. it does not have any effect \textcolor{red}{to the wave dynamics. Cancellation of the potential by rapid complex oscillation of its amplitude is a rather unique phenomenon of non-Hermitian dynamics which does not have any counterpart in Hermitian systems}.

 \section{Dynamics of a quantum particle in a rapidly-oscillating potential: effective potential description}
  Let us consider the problem of one-dimensional scattering/localization of matter or classical waves by a potential barrier/well $V(x)$ with a modulated ac amplitude $f(t)$. In dimensionless units, the dynamics is described by the Schr\"odiger-like wave equation for the wave function $\psi(x,t)$
  \begin{equation}
  i \frac{\partial \psi}{\partial t}=-\frac{\partial^2 \psi}{\partial x^2}+f(t) V(x) \psi
  \end{equation}
where $V(x) \rightarrow 0$ sufficiently fast as $|x| \rightarrow \pm \infty$ and $f(t)$ is a rapidly oscillating function of time with zero mean. In the standard description of wave scattering and localization by a rapidly oscillating potential \cite{r22,r24,r25,r26,r27,r27bis}, the amplitude $f(t)$ is assumed to be a periodic function of time, oscillating at a high-frequency $\omega$, and an effective time-independent Hamiltonian is obtained in power series expansion of $ \sim 1 / \omega$. The leading-order expansion term, valid up to the order $\sim 1 /\omega^2$, yields an effective static potential term, which is responsible for the most important dynamical effects like Kapitza stabilization \cite{r22,r24,r25,r26,r27,r27bis}. This is a rather counterintuitive result since in a naive description of the dynamics one would expect the rapidly oscillating potential to cancel out: since the particle is unable to follow the rapid variations of the potential, then it is likely to respond to its average value. In an Hemitian system, this is not the case because, even if the particle can not follow the rapid oscillations of the potential, these  are still able to affect the average dynamics by means of an effective -albeit small- potential contribution. Following Refs.\cite{r22,r27,r27bis}, the effective potential term can be derived in a rather simple way by means of the gauge transformation $\psi(x,t)=\phi(x,t) \exp [-i V(x) \int^t dt \xi f(\xi)]$, so that Eq.(1) yields
\begin{equation}
i \frac{\partial \phi}{\partial t}=- \left( \frac{\partial}{\partial x} - i \frac{\partial V}{\partial x} \int^t d \xi f(\xi) \right)^2 \phi
\end{equation}
and $\int^t d \xi f(\xi)$ is taken to have a vanishing mean value.
For a rapidly oscillating function $f(t)$, the leading-order approximate dynamics is obtained after averaging in time the right hand side of Eq.(2), disregarding the rapidly oscillating terms with zero mean. This yields the effective Schr\"odinger equation
\begin{equation}
i \frac{\partial \phi}{\partial t}= -\frac{\partial^2 \phi}{\partial \textcolor{red}{x}^2}+V_{eff}(x) \phi
\end{equation}
where the effective stationary potential $V_{eff}(x)$ is given by
\begin{equation}
V_{eff}(x)= \left( \frac{\partial V}{\partial x} \right)^2 \langle \left( \int^t d \xi f( \xi) \right)^2\rangle
\end{equation}
and where $\langle .. \rangle$ indicates time average over one oscillation cycle. \textcolor{red}{Note that the effective potential description holds for a rather general rapidly-oscillating ac field, which can be periodic, aperiodic or even stochastic} \footnote{\textcolor{red}{Kapitza-like stabilization of the inverted pendulum under quasi-periodic or stochastic driving is discussed, for instance, in:  Bogdanoff J.,  {\it J. Acoust. Soc. Am.}, {\bf  34} (1962) 1055; Dettmann C.P., Keating J.P. \and Prado S.D., {\it J. Phys. A}, {\bf 37} (2004) L377;  Simons Y.B., {\it Phys. Rev. E}, {\bf 80} (2009) 042102.}}. 
\textcolor{red}{In the most common case of a periodic modulation}, after introduction of the Fourier expansion $f(t)=\sum_n f_n \exp(i n \omega t)$, the time average term in the effective potential $V_{eff}$ reads explicitly
\begin{equation}
\langle \left( \int ^t d \xi f( \xi) \right)^2\rangle= \frac{1}{\omega^2}\sum_{n \neq 0} \frac{f_{-n} f_n}{n^2}
\end{equation}
For an ordinary Hermitian system, i.e. $f(t)$ and $V(x)$ real functions, the effective potential  $V_{eff}$ is always a non-vanishing one since $f_{-n}=f_n^*$ and the time average term (5) does not vanish. Albeit for a high oscillation frequency the effective potential is small  -it scales like $\sim 1 / \omega^2$ according to Eq.(5)- it can scatter waves of small momentum or can even sustain resonance modes, which are the signatures of Kapitza stabilization in the quantum realm \cite{r27}. The key point of our study is to consider a {\em complex} modulation function $f(t)$, which breaks the constraint $f_{-n}=f_n^*$. Interestingly, for a modulation function with a one-sided Fourier spectrum, e.g. for $f_n=0$ for $n \leq 0$, the time average term (5) vanishes, indicating that up to this order of approximation the effective potential is canceled. One might argue that, by considering higher-order correction terms in the asymptotic analysis (see e.g. \cite{r25,r26}) the effects of the rapidly-oscillating potential could be detected. However, for a one-sided Fourier spectrum of the modulation function $f(t)$ one can check that such additional terms (Eq.(16) of Ref.\cite{r25}) vanish as well, indicating that for such a wide class of complex modulation amplitudes the naive approach that would fully cancel the rapidly-oscillating potential is likely to be an exact result. Indeed, in the next section we prove that such a rapidly-oscillating potentials are scatteringless, i.e. they appear to be fully invisible, even in case of aperiodic oscillations.

\section{Universal scatteringless property of the non-Hermitian oscillating potential}
We are now going to prove the following general property of the Schr\"odinger equation (1) with an oscillating complex potential:\\ {\em Let $f(t)$ be an ac complex function with a one-sided Fourier spectrum $F(\omega)=(1 / 2 \pi) \int_{-\infty}^{\infty} dt f(t) \exp(i \omega)$ vanishing for $\omega> -\Omega_0$, with $\Omega_0>0$ possibly large for a rapidly-oscillating potential. Then any incident wave with frequency (energy) $\omega_0 \leq \Omega_0$ is not scattered off by the oscillating potential, which appears to be fully invisible.}
\par
 To prove the above theorem, let us indicate by $\psi_0(x,t)= \exp( ik_0x-i \omega_0 t)$ the free-particle plane wave solution to the Schr\"odinger equation (1) in the absence of the oscillating potential, i.e. for $f=0$, where $\omega_0=k_0^2$ is the frequency (energy) of the plane wave. For the sake of definiteness, let us assume $k_0>0$, corresponding to a fowardrd-propagating wave, however a similar analysis holds for a backward-propagating wave $k_0<0$. In the presence of the oscillating potential, let us look for a solution to Eq.(1) of the form
 \begin{equation}
 \psi(x,t)=\psi_0(x,t)+\int_{-\infty}^{\infty} d \omega \Theta (x, \omega) \exp(-i \omega t)
 \end{equation}
where the second term on the right hand side of Eq.(6), $\psi_1(x,t) \equiv \int d \omega \Theta(x, \omega) \exp(-i \omega t)$, accounts for the correction to the free-particle solution $\psi_0(x,t)$ induced by the scattering potential.
Substitution the Ansatz (6) into Eq.(1) yields the following non-homogeneous integro-differential equation for $\Theta(x, \omega)$ 
\begin{eqnarray}
& & \omega \Theta(x,\omega)+ \frac{\partial^2 \Theta}{\partial x^2} - V(x) \int d \Omega F(\omega-\Omega) \Theta(x, \Omega) \nonumber \\
& = & V(x) \exp(ik_0 x) F(\omega-\omega_0) 
\end{eqnarray}
 Note that, since $V(x) \rightarrow 0$ as $| x| \rightarrow \infty$, from Eq.(7) one has $ ( \partial^2 \Theta/ \partial x^2)+ \omega \Theta \sim 0$ as $| x| \rightarrow \infty$. Since we consider a forward-propagating incidence wave, the following asymptotic behavior of $\Theta(x, \omega)$ is assumed as $|x| \rightarrow \infty$
\begin{eqnarray}
\Theta(x, \omega) \sim \left\{
\begin{array}{ll}
r(\omega, \omega_0) \exp(-ikx) & x \rightarrow -\infty \\
 \left[ t( \omega,\omega_0) -\delta( \omega-\omega_0) \right] \exp( i k x) & x \rightarrow \infty 
\end{array}
\right.
\end{eqnarray}

\begin{figure}
\onefigure[width=7.8cm]{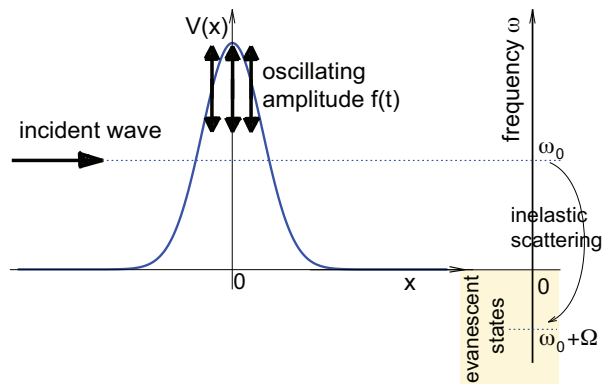}
\caption{(Color online) Schematic of the inelastic scattering of a plane wave at frequency $\omega_0$ from a bound potential $V(x)$ with an ac oscillating amplitude $f(t)$. A frequency component $\Omega$ of the ac oscillation induces a transition to a state with frequency $\omega=\omega_0+ \Omega$. If the Fourier spectrum $F(\Omega)$ of $f(t)$ vanishes for $\Omega>-\Omega_0$ and $\Omega_0 > \omega_0$, all inelastic scattered waves have negative frequencies, corresponding to evanescent (non-propagative) states.}
\end{figure}

\begin{figure*}
\onefigure[width=18cm]{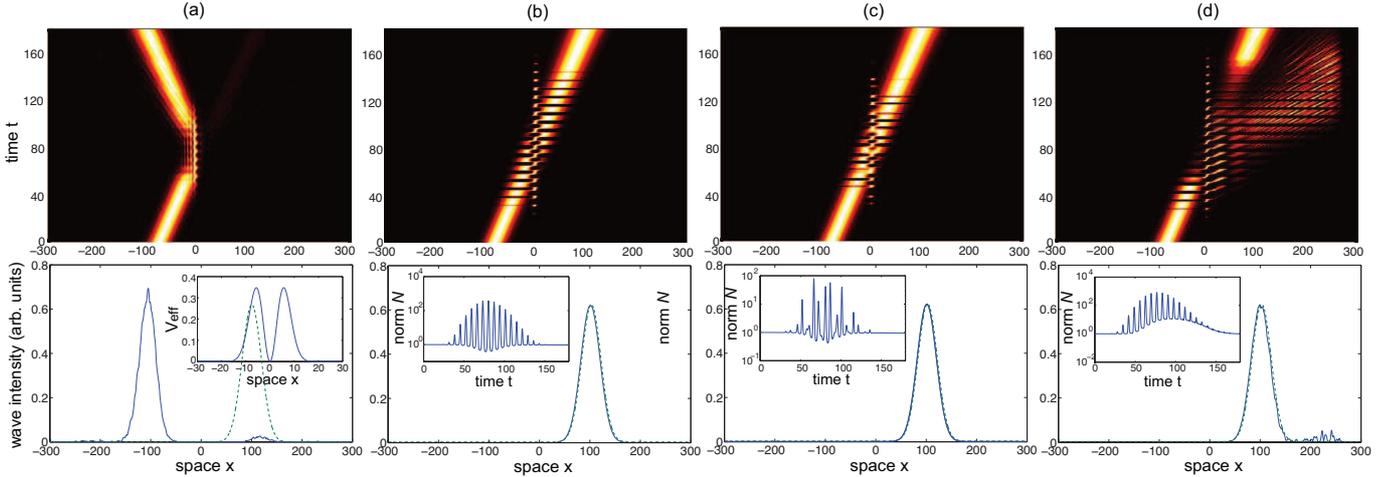}
\caption{(Color online) Scattering of a wave packet by the Gaussian potential (9) for four different modulation amplitudes $f(t)$: (a) $f(t)=\cos (\omega t)$, (b) $f(t)=0.5 \exp(i \omega t)$, (c) $f(t)=0.25 \exp(i \omega t)+0.25 \exp(i \sqrt{2} \omega t)$, and (d) $f(t)=0.5\exp(-i \omega t)$ with $\omega=0.9$, $V_0=7$ and $\beta=1/64$. 
 Other parameter values are given in the text. The upper panels show on a pseudocolor map the temporal evolution of the normalized wave intensity $\Psi(x,t)=|\psi(x,t)|^2 / {\rm max}_x |\psi(x,t)|^2$, whereas the lower panels show the wave intensity distribution $|\psi(x,t)|^2$ of the wave packet at time $t=180$ (solid curves). The dashed curves indicate, for comparison, the wave intensity distribution at the same time $t=180$ that one would observe in the absence of the scattering potential. In (b) and (c) the solid and dashed curves are almost overlapped, indicating that the oscillating potential is invisible. The inset in the lower panel (a) shows the shape of the effective potential $V_{eff}(x)$ in the high-frequency limit, given by Eq.(10). The insets in the lower panels of (b), (c) and (d) show on a log vertical scale the temporal evolution of the norm $\mathcal{N}(t)= \int dx |\psi(x,t)|^2$.}
\end{figure*}

where $k= \sqrt{\omega}>0$ for $\omega \geq 0$, and $k=i \sqrt{|\omega|}$ for $\omega<0$. Note that for $\omega>0$ the coefficients $r(\omega, \omega_0)$ and $t (\omega, \omega_0)$ represent the spectral reflection and transmission coefficients of scattered (propagative) plane waves \footnote{Note that the presence of reflected and transmitted waves at frequencies $\omega$ different than the frequency $\omega_0$ of the incident wave arises because of  inelastic scattering from the time-varying oscillating potential (see Fig.1).},   whereas for $\omega<0$ they are the amplitudes of evanescent (exponentially-damped) waves localized near the potential barrier/well. Since $F(\omega)$ vanishes for $ \omega> - \Omega_0$, from Eq.(7) it follows that the spectral amplitude $\Theta(x, \omega)$ depends on the other amplitudes $\Theta(x, \Omega)$ with $\Omega> \omega+\Omega_0$ solely. Let us assume $\omega_0 \leq \Omega_0$, i.e. the frequency (energy) of the incident wave below the cut-off frequency $\Omega_0$ of the spectrum $F(\omega)$. In this case the driving (inhomogeneous) term on the right hand side of Eq.(7) vanishes for any $\omega>0$. Therefore, for $\omega_0 \leq \Omega_0$ it follows that the solution $\Theta(x,\omega)$ to Eq.(7) vanishes for any $\omega > 0$ \footnote{More precisely one has $\Theta(x, \omega)=0$ for any $\omega >\omega_0-\Omega_0$.}. This means that, provided that $\omega_0 \leq\Omega_0$, the correction $\psi_1(x,t)$ to the free-particle state $\psi_0(x,t)$ introduced by the oscillating potential 
is composed solely by {\it evanescent} waves, which exponentially decay far from the potential region. As a consequence, from Eq.(8) one has $r(\omega, \omega_0)=0$ and $t(\omega, \omega_0)=\delta(\omega-\omega_0)$ for any $\omega >0$, i.e. the scattering potential is fully invisible and does not scatter the incident plane wave at frequency $\omega_0$, provided that $\omega_0 \leq \Omega_0$. Note that such a property holds for any oscillating amplitude $f(t)$ with a one-sided spectrum $F(\omega)$, i.e. it is not necessarily a periodic function. Note also that, for a rapidly oscillating potential, i.e. $\Omega_0$ large, the  spectral invisibility region $(0, \Omega_0)$ of the potential can be made arbitrarily large.\\
From a physical viewpoint, such an invisibility effect can be explained by considering the oscillating potential $f(t) V(x)$ in Eq.(1) as a time-dependent perturbation to the free-particle Schr\"odinger equation. The effect of the perturbation is to scatter the incident plane wave $\psi_0(x,t)$ at frequency $\omega_0$, i.e. to induce transitions into the other energy states. Since $f(t)$ has a one-sided Fourier spectrum $F(\Omega)$ with non-vanishing frequency components $\Omega<-\Omega_0$, owing to energy conservation the perturbation can induce transitions to states with frequency $\omega=\omega_0+ \Omega <\omega_0-\Omega_0$. For $\omega_0< \Omega_0$, this implies that the only allowed transitions are those toward waves with negative frequencies (see Fig.1), which are necessarily evanescent waves. Therefore the potential is not able to scatter the incident plane wave. Such a transparency effect, corresponding to a flat (unit) spectral transmission coefficient of plane waves, indicates that the oscillating potential is unlikely to sustain quasi-bound (resonance) states, which  would correspond to sharp changes of the spectral transmission near the resonance.\\
As a final comment it should be noted that, when the  complex ac oscillation function $f(t)$ has a one-sided Fourier spectrum $F(\omega)$ vanishing for $\omega>\Omega_0$ (rather than for $\omega<- \Omega_0$ as in the theorem stated above), the oscillating potential does not turn out to be invisible, even though the effective averaged potential (4) vanishes. In fact, in this case following the lines of the demonstration of the theorem given above one can conclude that $\Theta(x,\omega)=0$ for $\omega< \omega_0+ \Omega_0$, i.e. inelastic scattering to high-frequency and fast propagative waves is now allowed. While the contribution to the scattered field introduced by such a high-frequency components can be small, it makes the oscillating bound potential visible.

\section{An example: scattering and localization from an oscillating Gaussian potential}
To check and exemplify the theoretical predictions presented in the previous sections, let us consider the dynamical stabilization and scattering properties of a Gaussian potential
\begin{equation}
V(x)=V_0 \exp(- \beta x^2)
\end{equation}
under a rapidly oscillation ac amplitude $f(t)$. In the Hermitian limit, the Gaussian potential was previously considered in Refs.\cite{r25,r27} as an example of the quantum Kapitza stabilization. In fact, for a sinusoidal oscillation amplitude $f(t)= \cos (\omega t)$, in the high-frequency oscillation regime the effective potential $V_{eff}(x)$ [Eq.(4)]
\begin{equation}
V_{eff}(x)=\frac{2 V_0^2 \beta^2}{\omega^2} x^2 \exp(-2 \beta x^2)
\end{equation}
is a double-humped potential [see the inset if the lower plot of Fig.2(a)] that can sustain quasi-bound modes \cite{r27}.\\ 
We numerically studied the scattering and localization properties of the oscillating Gaussian potential by considering three different ac oscillation amplitudes:\\ (i) the sinusoidal oscillation 
\begin{equation}
f(t)=\cos(\omega t);
\end{equation} 
(ii) the periodic non-Hermitian oscillation 
\begin{equation}
f(t)=\frac{1}{2} \exp( i \omega t);
\end{equation} 
(iii) the quasi-periodic non-Hermitian oscillation 
\begin{equation}
f(t)=\frac{1}{4} \left[ \exp(i \omega t)+ \exp( i \sqrt{2} \omega t) \right] .
\end{equation} 
The first case corresponds to the Hermitian problem with the effective potential given by Eq.(10), whereas the latter two cases correspond to a vanishing effective potential.\\ 
In a first set of simulations, we numerically studied the scattering properties of the oscillating potential by considering a spatially-wide (spectrally-narrow) Gaussian wave packet  with a small carrier wave number $k_0$ incident onto the scattering potential from the left side. 
Figure 2 shows the numerically-computed evolution in time of the wave packet intensity $|\psi(x,t)|^2$, normalized to its peak value at any time $t$, i.e. of the distribution $\Psi(x,t)=|\psi(x,t)|^2 / {\rm max}_x |\psi(x,t)|^2$, for an initial field distribution, at time $t=0$, given by $\psi(x,0) \propto \exp[-(x+d)^2/w_0^2+ik_0x]$, with $d=80$, $w_0=25$ and $k_0=0.5$. The detailed behavior of the wave intensity $|\psi(x,t)|^2$, at the time $t=180$ after traversing the oscillating potential, is shown in the lower plots of the figure and compared with the field intensity that one would observe in the absence of any potential (free particle wave packet propagation). In the Hermitian case [Fig.2(a)], the wave packet is clearly scattered off by the oscillating potential. The scattering arises from the non-vanishing effective potential $V_{eff}(x)$: in fact, the carrier frequency $\omega_0=k_0^2=0.25$ of the  incident wave packet is below the double-humped barrier height of the effective potential [see the inset in the lower plot of Fig.2(a)]. Conversely, in the periodic and quasi-periodic non-Hermitian oscillations defined by Eqs.(12) and (13) there is not any reflected wave packet. By comparing the field distributions after the interaction with the oscillating potential  [lower plots in Figs.2(b) and (c)], the wave packet after crossing the oscillating potential is propagated as if there were no potential at all, demonstrating the complete invisibility of the potential.
Noe that, while in the Hermitian case (i) the norm $\mathcal{N}(t)=\int dx  |\psi(x,t)|^2$ is conserved, in the non-Hermitian cases (ii) and (iii) $\mathcal{N}$ is not conserved and undergoes large oscillations following the modulation cycles of $f(t)$ [see the insets in the lower plots of Figs.2(b) and (c)]. We also checked that, according to the theoretical analysis presented in the previous section, the complex ac oscillation with one-sided Fourier spectrum in the negative-frequency range is not invisible. Figure 2(d) shows, as an example, the numerically-computed wave packet evolution for an oscillation function $f(t)=(1/2) \exp(-i \omega t)$. i.e. with reversed sign of oscillation frequency as compared to Fig.2(b). Note that in this case fast propagating wave components are clearly generated when the wave packet crosses the potential scattering region, which make the potential not an invisible one.  
\par
In a second set of simulations we investigated the dynamical (Kapitza) stabilization properties of the oscillating Gaussian potential (9). In this case as an initial condition we assumed a narrower Gaussian-shaped wave packet with zero mean wave number, namely we assumed $\psi(x,0) \propto \exp(-x^2/w_0^2)$ with $w_0=5$. The numerical results of wave packet propagation are sown in Fig.3. In the Hermitian modulation case $f(t)=\cos(\omega t)$ [Fig.3(b)], the double-humped effective potential (10) clearly provides dynamical localization of the wave packet, greatly reducing its spreading as compared to the free-evolution regime [Fig.3(a)]. Conversely, for the non-Hermitian modulations (12) and (13) [Figs.3(c) and (d)] the wave packet undergoes a breathing dynamics in each oscillation cycle of the modulation, however on average the wave packet spreads like in the free-particle regime. The breathing dynamics arises because of the alternating amplification/ attenuation of the wave intensity, in each oscillation cycle, near the potential region $x=0$, corresponding to an effective localization/delocalization of the wave packet at around $x=0$. The detailed temporal evolution of the wave packet widths $w(t)= \sqrt{\int dx x^2 |\psi(x,t)|^2 / \int dx |\psi(x,t)|^2}$ in the different modulation regimes is shown in Fig.3(e). Note that in the dynamical (Kapitza)  stabilization regime (curve b) the wave packet width $w(t)$ shows a residual increase with time $t$, but at a rate smaller than the one observed in the free-particle case (curve a). The residual increase of the wave packet width in time arises because of the leaky nature of the double-humped effective potential (10), which does not sustain truly bound states but resonance modes.    

\textcolor{red}{Finally, let us briefly mention possible physical implementations of non-Hermitian oscillating potentials, where cancellation of dynamical stabilization and scattering could be experimentally observed. A first platform is provided by electromagnetic or light propagation in engineered dielectric media, where the refractive index profile of the medium plays the same role as the scattering potential in the Schr\"odinger equation and the temporal evolution of the wave function corresponds to the spatial propagation  of the light field along the axial direction of the medium \cite{r56}. By periodically varying the refractive index distribution along the propagation distance \cite{r34,r35,r36}, one could in principle be able to observe cancellation of dynamical localization, i.e. of light guiding, when spatial profiles of real and imaginary parts of the refractive index satisfy the conditions discussed above. We note that refractive index engineering, capable of controlling the real (index) and imaginary (absorption/gain) properties in graded-index media, has  been recently demonstrated in Ref.\cite{r57} at microwaves. Other physical platforms could be envisaged by considering the discrete version of the Schr\"odinger equation (1). In fact, Kapitza stabilization can be observed for discrete waves in oscillating potentials  on a lattice as well \cite{r33}. Chains of coupled optical waveguides \cite{r33}, coupled microring resonators  or coupled optomechanic oscillators \cite{r54,r58} with modulated parameters could provide other physical systems to observe the discrete version of scatteringless potentials.} 
 
\begin{figure}
\onefigure[width=8.6cm]{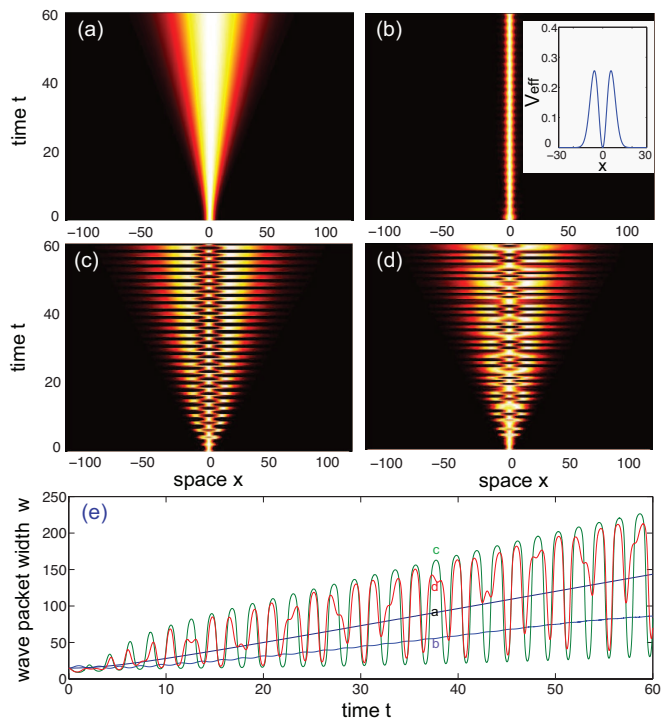}
\caption{(Color online) Localization properties of the oscillating Gaussian potential (9). Panels (a-d) show the numerically-computed temporal evolution, on a pseudo color map, of the normalized wave intensity  $\Psi(x,t)=|\psi(x,t)|^2 / {\rm max}_x |\psi(x,t)|^2$  (a) in the absence of the potential (free-propagation wave packet spreading); and for three different modulation amplitudes $f(t)$: (b) $f(t)=\cos (\omega t)$, (c) $f(t)=0.5 \exp(i \omega t)$, and (d) $f(t)=0.25 \exp(i \omega t)+0.25 \exp(i \sqrt{2} \omega t)$ with $\omega=3$, $V_0=20$ and $\beta=1/64$. The inset in (b) shows the behavior of the effective potential (10) for the sinusoidal modulation case $f(t)=\cos(\omega t)$. The temporal evolution of the wave packet width $w(t)$ is shown in (e). Curves a,b,c, and d refer to the wave packets shown in (a), (b), (c) and (d), respectively.}
\end{figure}

\section{Conclusions} Rapid modulations in classical and quantum Hamiltonian systems are known to modify the system dynamics in a nontrivial way, because the high-frequency limit of the Floquet Hamiltonian of the periodically-driven system is not equal to the time-averaged Hamiltonian. A paradigmatic example of nontrivial dynamics induced by a rapid modulation of the potential is the dynamical (Kapitza) stabilization effect, i.e. the possibility for a classical or quantum particle to be trapped by a rapidly oscillating potential in cases where the static potential cannot trap them. Even though the particle is not able to follow the rapid external oscillations of the potential, these  are still able to affect the average dynamics by means of an effective -albeit small- potential contribution, which can induce particle trapping. An intriguing question is whether one can find cases where the rapidly oscillating potential does not provide any effect on the particle dynamics, i.e. where the rapid oscillations effectively cancel the potential. Such a kind of invisible potentials are prevented in ordinary Hermitian quantum mechanics, because the effective potential describing the leading-order asymptotic dynamics is always a non-vanishing potential \cite{r22,r35,r26,r27}: it is able to scatter particles with small momentum or to trap them.  In this work we considered the scattering and dynamical stabilization properties of bound potentials $V(x)$ with an ac oscillating {\em complex} amplitude $f(t)$ and predicted that for a wide class of  modulation functions possessing a one-sided Fourier spectrum the oscillating potential can be effectively canceled. \textcolor{red}{Our results shed new light into the physics of driven dynamical systems and dynamical stabilization, indicating that extension of potential modulation into the non-Hermitian realm enables a more flexible control of the scattering and stabilization properties of classical or matter waves. For example, by switching the modulation amplitude of the potential from purely real (Hermitian) to mixed real-imaginary (non-Hermitian) one could in principle dynamically change the trapping properties of a system, with potential applications to dynamical control of the wave guiding  properties of light and invisibility of dielectric media \cite{r49,r57}.}

\end{document}